\documentclass[a4paper]{article}
\newcommand{\Rset}{\Re} 
\newtheorem{conject}{Conjecture}
\newtheorem{axiom}{Axiom}

\newcommand{\M}{{\bf M}}   

\newcommand{\A}{\mbox{${\cal A}$}} 
\newcommand{\BB}{\mbox{${\cal B}$}}
\newcommand{\MM}{{\cal M}} 
\newcommand{\XX}{{\cal X}} 
\newcommand{\YY}{{\cal Y}} 
\newcommand{\PP}{{\cal P}}
\newcommand{\QQ}{{\cal Q}}

\newcommand{\x}{{\em x}}
\newcommand{\y}{{\em y}}

\newcommand{\eg}{{\em eg.}}
\newcommand{\etc}{{\em etc.}}
\newcommand{\ie}{{\em i.e.}}
\newcommand{\viceversa}{{\em vice versa}}

\newcommand{\To}{0}
\newcommand{\Ti}{I}

\begin{document}
\title{The Logic of Quantum Mechanics Derived from Classical 
General Relativity}
\author{Mark J Hadley\footnotemark \\ Dept. of Physics, Univ. of
Warwick\\  COVENTRY CV4~7AL, UK}
\maketitle
\begin{abstract}
For the first time it is shown that the logic of quantum mechanics can
be derived from Classical Physics. An orthomodular lattice of
propositions, characteristic of quantum logic, is constructed for
manifolds in Einstein's theory of general relativity. A particle is
modelled by a topologically non-trivial 4-manifold with closed
timelike curves - a 4-geon, rather than as an evolving 3-manifold. It
is then possible for both the state preparation {\em and} measurement
apparatus to constrain the results of experiments. It is shown that
propositions about the results of measurements can satisfy a
non-distributive logic rather than the Boolean logic of classical
systems.  Reasonable assumptions about the role of the measurement
apparatus leads to an orthomodular lattice of propositions
characteristic of quantum logic.
\end{abstract}

\bibliographystyle{unsrt}
\section{Comment}\renewcommand{\thefootnote}{\fnsymbol{footnote}}
This paper has been published in Foundations of Physics Letters
\cite{hadley97}. The work forms the basis of my doctoral thesis
\cite{hadley_thesis}. A short, less formal talk about my work is
archived in quant-ph/9609021 \cite{hadley}\footnotetext[1]{email:
M.J.Hadley@warwick.ac.uk}. 
\section{INTRODUCTION}
Quantum logic is characterised by the propositions of an orthomodular
lattice, the distinguishing feature of which is the failure of the
distributive law which is replaced with the weaker orthomodularity
condition\cite{beltrametti_cassinelli}. From this orthomodular lattice
it is then possible to generate the Hilbert space structure of quantum
mechanics\cite{holland_s}. That quantum systems satisfy a non-Boolean
logic is an experimental fact that has never been satisfactorily
explained.

This paper suggests an origin for quantum logic and formally
constructs an orthomodular lattice of propositions about manifolds
in classical general relativity.   

Since the formulation of general relativity,
attempts\cite{einstein_rosen,misner_wheeler} have been made to model
elementary particles as topologically non-trivial structures in
spacetime called geons. These models exhibited interesting
particle-like properties, such as mass and charge without apparent
sources, but could not reproduce the features of quantum mechanics
which the particles must obey. These models were, however, based on
{\em three}-manifolds evolving in time, in the sense that there was a
global hypersurface permitting the definition of a global time
coordinate. Once the manifold was defined at a given time its
evolution was deterministic - independent of subsequent measurements
that may, or may not, be made.

There has recently been much speculation about the existence of closed
timelike curves (CTCs), their stability and the possibility of time
travel with its associated inconsistencies
\cite{friedman_morris}. Although CTCs appear unphysical, there is
nothing in the theory of general relativity to exclude
them\cite{thorne}. General relativity treats spacetime as a manifold,
which is locally diffeomorphic to $\Rset^4$, but does not prescribe
its topology\cite{misner_wheeler}. Interacting classical objects (or
fields) in a spacetime with CTCs require additional boundary
conditions to uniquely determine their evolution
\cite{friedman_morris}. Data that would normally be sufficient to
define a unique trajectory may, in the presence of CTCs, permit more
than one possibility.

This work suggests a model for a particle as a non-trivial topological
structure in four dimensions (space and time) not just space. While it
is an obvious extension of the old ideas on geons, it makes fuller use
of the richness of general relativity than did the earlier work. Thus
the extension from topologically non-trivial 3-spaces (3-geons) to
spacetime (4-geons) gives rise to the possibility of CTCs and the
associated {\em impossibility\/} of defining a global time coordinate.

It is well known that if states are modelled as projections of a
complex Hilbert space then the symmetries of the spacetime (Galilean
or Poincar\'{e}) together with the appropriate internal symmetries of
the object leads inevitably to the familiar equations of
non-relativistic and relativistic quantum mechanics respectively,
together with commutation relations and a universal constant with the
dimensions of Planck's constant\cite{ballentine}.

The conjectured 4-geon description of particles is speculative but
this single assumption is able to
unify the particle and field descriptions of nature, explain quantum
logic and in doing so reconciles general relativity and quantum
mechanics.
\setcounter{footnote}{2}

\section{4-GEON}
The present analysis is based upon a model of an elementary particle
as a distortion of spacetime, (a four dimensional semi-Riemannian
manifold with non-trivial topology). The manifold includes both the
particle and the background metric, and being four dimensional without
a {\em global} time coordinate, the particle and its evolution are
inseparable - they are both described by the four-manifold. We now
express the properties, which we require of a particle, in the
language of manifolds.

\begin{axiom}[Asymptotic flatness] Far away from the particle spacetime
is topologically trivial and asymptotically flat with an
approximately Lorentzian metric. 
\label{ax:flat}
\end{axiom}
In mathematical terms - spacetime is a 4-manifold, $\MM$ and there
exists a 4-manifold K, such that $\MM / K $ is diffeomorphic to
$\Rset^4/({\bf B}^3 \times \Rset)$ and the metric on $\MM / K $ is
asymptotically Lorentzian.\footnote{${\bf B}^3$ is a solid sphere} $K$
or (${\bf B}^3 \times \Rset$) can be regarded as the world-tube within
which the `particle' is considered to exist.

This axiom formally states the fact that we experience an
approximately Lorentzian spacetime, and that if space and time are
strongly distorted and convoluted to form a particle then that region
can be localised. (It may be noted that asymptotic flatness is not a
reasonable property to require for a quark because it cannot be
isolated [there is no evidence of an isolated quark embedded in a flat
spacetime] therefore the present work cannot be applied automatically
to an isolated quark.)

The position of a distortion of spacetime is not a trivial concept -
it implies a mapping from the 4-manifold, which is both the particle
and the background spacetime, onto the flat spacetime used within the
laboratory. There is in general no such map that can be defined
globally, yet a local map obviously cannot relate the relative
positions of distant objects. This axiom gives a practical definition
of the position of a particle - it is the region where the non-trivial
topology resides. Any experimental arrangement which confines (with
barriers of some sort) the ${\bf B}^3$ region of non-trivial topology,
defines the position of the particle. From this axiom, the region
outside the barrier is topologically trivial and therefore {\em does}
admit global coordinates.

Using the asymptotic flatness axiom it is now possible to define what is
meant by a particle-like solution:
\begin{axiom}[Particle-like] In any volume of 3-space an
experiment to determine the presence of the particle will yield a
true or a false value only.
\label{ax:particle}
\end{axiom}
This is consistent with the non-relativistic indivisibility of the
particle. By contrast, a gravitational wave may be a diffuse object
with a density in different regions of space which can take on a
continuous range of values. An object which did not satisfy this axiom
(at least in the non-relativistic approximation) would not be
considered to be a particle. The axiom is clearly satisfied by
classical particles and, because it refers only to the result of a
position measurement, it conforms also with a quantum mechanical
description of a particle.

The particle-like axiom requires the property of asymptotic flatness,
defined above, to give meaning to a 3-space. The three space is
defined in the global asymptotically flat, topologically trivial
region, $\MM / K$, which is diffeomorphic to $\Rset^4/({\bf B}^3
\times \Rset)$ as defined above. 

We are now able to state the required properties of a 4-geon.
\begin{conject}[4-Geon] A particle is a semi-Riemannian 
spacetime manifold, $\MM$, which is a solution of Einstein's equations
of general relativity. The manifold is topologically non-trivial, with
a non-trivial causal structure, and is asymptotically flat and
particle-like (Axiom~\ref{ax:particle}).
\label{conj:4geon}
\end{conject}
It would be very appealing if $\MM$ was a solution of the vacuum
equations\cite{einstein_rosen}, but for the arguments that follow this
is not essential; unspecified non-gravitational sources could be part
of the structure. The assumed existence of CTCs as an integral part of
the structure (rather than as a passive feature of the background
topology) is an essential feature of the manifold; when they exist
additional boundary conditions may be required to define the
manifold\cite{friedman_morris}. This aspect of the structure provides
a causal link between measurement apparatus and state preparation,
permitting {\em both} to form part of the boundary conditions which
constrain the field equations.

The axioms formally state conditions that any description of a
particle must reasonably be expected to satisfy. In contrast, the
4-geon (Conjecture~\ref{conj:4geon} above) is novel and speculative
since it is not known whether such solutions exist - either to the
vacuum or the full field equations of general relativity; however,
{\em there no reason to suppose that they cannot exist}. It will be
shown that this single speculative element not only yields quantum
logic, but is sufficient to derive the equations of quantum mechanics
and in doing so reconciles general relativity with quantum
mechanics. Although this work proposes novel and unproven structures
in general relativity it requires neither a modification, nor any
addition to Einstein's equations; the number of spacetime dimensions
remains 3~+~1. The work does not require extraneous particle fields
(as used in conventional quantum field theory), nor does it impose a
quantum field of unknown origin (as does Bohm's theory). In short,
this single conjecture is sufficient to unify particle and
gravitational field descriptions of Nature, quantum and classical
logic and quantum mechanics with general relativity.
 
\section{STATE-PREPARATION AND MEASUREMENT}

The role of both the measurement and state preparation in defining the
4-geon is crucial. It is self-evident that state preparation sets
boundary conditions. Whether we regard a particle as a classical
billiard ball, a quantum of a quantum field, or a classical field, the
state preparation limits the possibilities; it restricts the possible
solutions to those consistent with the apparatus. Systems with slits,
collimators and shutters provide obvious boundary conditions which any
solution must satisfy. For a geon, or a 4-geon, a barrier is a region
which the topologically non-trivial region cannot traverse. Such
barriers can be used to form slits and collimators \etc. and they
obviously restrict the space of possible solutions. We state this
formally as an axiom:

\begin{axiom}[State preparation] The state preparation sets boundary
conditions for the solutions to the field equations.
\label{ax:state}
\end{axiom}
The exact nature of these boundary conditions,
and whether they can always be equated with physical barriers
such as collimators, is irrelevant to the analysis that
follows. 

Consider now an apparatus associated with a measurement, which is in
many respects similar to that involved in a state
preparation. Arrangements of slits, barriers and collimators are
common features of the measurement apparatus. They are constructed
from barriers which cannot be traversed by the non-trivial topology,
which is the particle.
 
We take as a paradigm for a position measurement that barriers divide
space into regions which are then probed (in any manner) to ascertain
the existence, or otherwise, of the particle in a region. The
particle-like axiom and the asymptotic flatness axiom assures us that
the topologically non-trivial region can be confined but not split.

We take the view of Holland \cite{holland} that most measurements can be
reduced to position measurements. A sequence of shutters and
collimators and filters (\eg such as used in a Stern-Gerlach
apparatus) determines the state preparation, while a very similar
system of shutters \etc, resulting in confinement to one of a number
of regions and subsequent detection, acts as a measurement.

For a classical object there is no causal connection that could allow
the measurement conditions to influence the evolution. If the state
preparation was insufficient to uniquely specify the trajectory there
would be a statistical distribution of possible initial states, each
of which would evolve deterministically. By contrast on a spacetime
with CTCs extra conditions are required for a unique deterministic
evolution \cite{friedman_morris}. With a particle modelled as a 4-geon
however, there {\em would} be a causal link allowing the measurement
conditions to contribute to the definition of the 4-manifold. A 4-geon
is a 4-dimensional spacetime manifold which satisfies the boundary
conditions set by {\em both} the state preparation {\em and} the
measurement. This justifies a further axiom:

\begin{axiom}[Measurement process] The measurement process sets
boundary conditions for the 4-geon which are not necessarily
redundant, in the sense that they contribute to the definition of the
4-manifold.
\label{ax:meas}
\end{axiom}
This axiom is inevitable if the particle contains CTCs, because the
state preparation and the measurement conditions can no longer be
distinguished by causal arguments.

\begin{axiom}[Exclusive experiments] Some pairs of measurements are
mutually exclusive in the sense that they cannot be made simultaneously.
\end{axiom}
This axiom expresses an established experimental fact -
see\cite[Chapter~7]{Bohr}. The famous examples of two such
complementary variables are \x-position and \x-momentum. The \x\ and
\y\ components of spin form another pair of complementary variables
with a very simple logical structure. That measurements cannot be made
simultaneously is still consistent with classical physics; objects
would have a precise position and momentum, but we could only measure
one property or the other. Quantum mechanics goes much further and
asserts that a particle cannot {\em even} posses precise values of
both properties simultaneously. The present work is unique in
explaining why an inability to make simultaneous measurements should
lead to incompatible observables in the quantum mechanical sense.
 
\section{PROPOSITIONS AND 4-MANIFOLDS}
We now consider the semi-Riemannian manifolds, $\M$, that could
satisfy the different boundary conditions imposed by state preparation
and measurement:
Let $\MM \equiv \{\M\}$ denote the set of 4-manifolds consistent with
the state preparation conditions; there is no reason to
suppose that a $\M $ is unique. The inability to define $\M $
uniquely will result in a {\em classical} distribution of measurement
results.

\setlength{\unitlength}{0.8mm}
\begin{figure}[p]
\begin{picture}(220,100)(15,0)
\put(5,5){\framebox(170,90)[tl]{$\{\M\}$}}

\thicklines

\put(55,50){\oval(80,40)}
\put(15,70){\makebox(0,0)[br]{$\PP$}}
\put(55,30){\line(0,1){40} }
\put(25,50){\makebox(0,0)[l]{$\PP^-$}}
\put(85,50){\makebox(0,0)[r]{$\PP^+$}}

\put(140,50){\oval(40,80)}
\put(160,90){\makebox(0,0)[bl]{$\QQ$}}
\put(120,50){\line(1,0){40} }
\put(130,20){\makebox(0,0)[b]{${\bf \QQ^-} $}}
\put(130,80){\makebox(0,0)[t]{${\bf \QQ^+} $}}

\thinlines
\end{picture}
\caption{Sets of 4-manifolds consistent with both state preparation
and the boundary conditions imposed by different measurement
conditions.}
\label{fig:venn}
\end{figure}
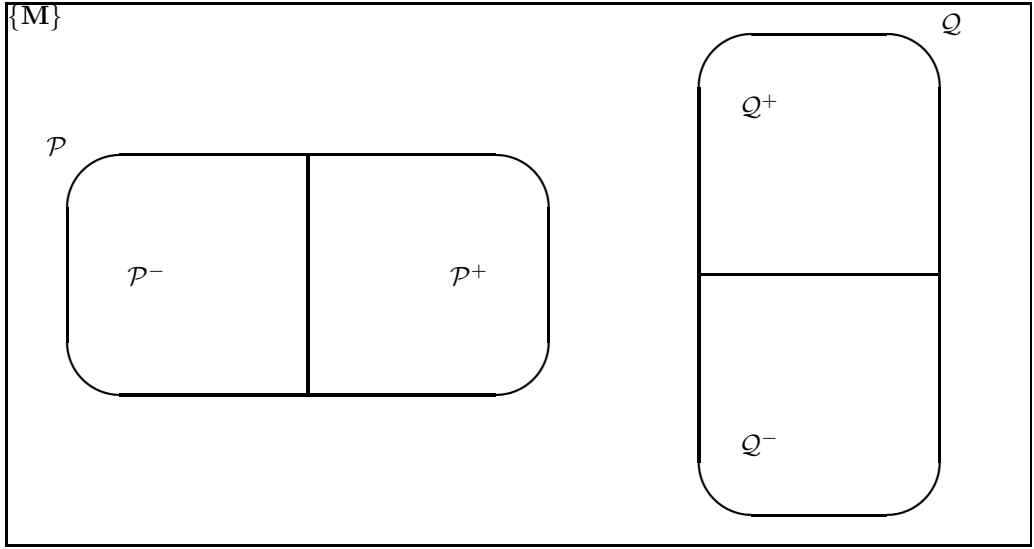

The 4-manifold describes both the particle and its evolution; for a
4-geon they are inseparable. Consequently, the terms {\em initial\/}
and {\em evolution\/} need to be used with great care. Although valid
in the asymptotically flat region (and hence to any observer), they
cannot be extended throughout the manifold. Preparation {\em followed
by\/} measurement is also a concept valid only in the asymptotic
region: {\em within the particle causal structure breaks down}.

Consider first the case of the classical 3-geon. Each $\M$ corresponds
to an evolving 3-manifold $\M^3$. Each $\M^3$ will evolve
deterministically in a way determined uniquely by the Einstein field
equations and the initial condition $\M^3(t_0)$ (the distribution of
$\M^3(t_0)$ determines the distribution of $\M^3(t)$ at any later time
$t_1 > t_0$). If the geon is particle-like, then any experiment that
depends upon a position measurement will give a result for each $\M^3$
at any time.  Consequently, the boundary conditions imposed by
measurements are necessarily compatible with any 3-geon that satisfies
the particle-like proposition; in other words they are redundant.

By contrast, the 4-geon with CTCs as part of its structure cannot be
decomposed into a three manifold and a time variable. It is known that
further boundary conditions need not be redundant in a spacetime which
admits CTCs\cite{friedman_morris}. In principle, the measurement apparatus
itself can provide additional boundary conditions.

Consider measurements P, Q for which the boundary conditions cannot be
simultaneously applied. They could be the \x-comp\-on\-ent of spin and
\y-comp\-on\-ent of spin, or \x-position and \x-momentum; for
simplicity we will consider two-valued measurements (\eg\ spin up or
down for a spin-half particle or \x-position $>0$ and \x-momentum $>
0$). We will denote the result that {\em ``the state has a +ve P
value''\/} by $P^+$, ($P^-,Q^+,Q^-$, are defined similarly). As
propositions, $P^+$ is clearly the complement of $P^-$; if $P^+$ is
true then $P^-$ is false and \viceversa, and similarly for $Q^+$ and
$Q^-$.

As before, let $\MM \equiv\{\M\}$ denote the set of 4-manifolds
consistent with the state preparation. The measurements define subsets
of $\{\M\}$; we denote by $\PP$ those manifolds consistent with the
state preparation {\em and} the boundary conditions imposed by a
P-measurement. $\PP$ is clearly the disjoint union of $\PP^+$ and
$\PP^-$ - the manifolds corresponding to $P^+$ and $P^-$,
respectively. Where the boundary conditions imposed by the measurement
are not redundant $\{\M\}$, $\PP$ and $\QQ$ need not be the same (see
Figure \ref{fig:venn}). There is a one-to-one correspondence between
the sets of manifolds in the Figure and the four non-trivial
propositions, $p,q,r,s$. However, two statements, or experimental
procedures correspond to the same proposition if they cannot be
distinguished by any state preparation - in other words if they give
exactly the same information about each and every state. Therefore the
statement that {P has a value} is always true by the particle-like
Axiom \ref{ax:particle}, as is the statement that {Q has a value;}
hence the subsets $\PP$ and $\QQ$ correspond to the {\em same}
proposition $\Ti$ and we have the possibility:
\begin{equation}
\PP^+ \neq (\PP^+ \cap \QQ^+)\cup(\PP^+ \cap \QQ^-)
\label{eq:noncl}
\end{equation}
If the boundary conditions are incompatible then $\PP$ and $\QQ$ are
disjoint and the following holds (see Figure~\ref{fig:venn}):
\begin{equation}
0 = (\PP^+ \cap \QQ^+) = (\PP^+ \cap \QQ^-) \neq \PP^+
\end{equation}
{\bf Therefore, propositions about a state do not necessarily
satisfy the distributive law of Boolean algebra.} 

\section{GENERAL RELATIVITY AND QUANTUM MECHANICS}
The significance of this result (Equation \ref{eq:noncl}) is that the
failure of the distributive law is synonymous with the existence of
incompatible observables\cite[Page~126]{beltrametti_cassinelli}; it is
a definitive property of non-classical systems of which a system
obeying the rules of quantum mechanics is an example. To obtain
quantum mechanics (as represented by a projections of a Hilbert space)
we need to replace the distributive law with the weaker orthomodular
condition:
\begin{equation}
a \le b \Rightarrow b = a \vee (b \wedge \hbox{NOT}(a))
\label{eq:quan}
\end{equation}

where $\le$ is a partial ordering relation which is transitive,
reflexive and antisymmetric; it corresponds to set theoretic
inclusion of the manifolds, $\A \subseteq \BB$. 

For propositions, $a$ and $b$, the ordering relation can only be
applied if they can be evaluated together
\cite[Chapter~13]{beltrametti_cassinelli}. When $a \le b$ there must
be a measurement apparatus which enables $a$ and $b$ to be measured
together. Let $\PP$ be the subset of $\MM$ defined by this measurement
(see Figure~\ref{fig:mod_venn}). Then $\A^+ \subseteq \BB^+ \subseteq
\PP$ and the complements with respect to $\PP$ satisfy $\BB^-
\subseteq \A^- \subseteq \PP$. Clearly:
\begin{equation}
\A^+ \subseteq \BB^+ \Rightarrow \BB^+ = \A^+ \cup (\BB^+ \cap  \A^-)
\end{equation}

\setlength{\unitlength}{0.8mm}
\begin{figure}[p]
\begin{picture}(220,100)(15,0)
\put(5,5){\framebox(170,90)[tl]{$\{\M\}$}}

\thicklines

\put(55,50){\oval(80,40)}
\put(15,70){\makebox(0,0)[br]{$\PP$}}
\put(55,30){\line(0,1){40} }
\put(68,43){\circle{20}}
\put(68,43){\makebox(0,0)[c]{$\A^+$}}
\put(25,50){\makebox(0,0)[l]{$\BB^-$}}
\put(85,50){\makebox(0,0)[r]{$\BB^+$}}

\put(140,50){\oval(40,80)}
\put(160,90){\makebox(0,0)[bl]{$\QQ$}}
\put(120,50){\line(1,0){40} }
\put(130,20){\makebox(0,0)[b]{${\bf \QQ^-} $}}
\put(130,80){\makebox(0,0)[t]{${\bf \QQ^+} $}}

\thinlines
\end{picture}
\caption{Sets of 4-manifolds illustrating the orthomodular condition
for compatible propositions, $a$ and $b$.}
\label{fig:mod_venn}
\end{figure}
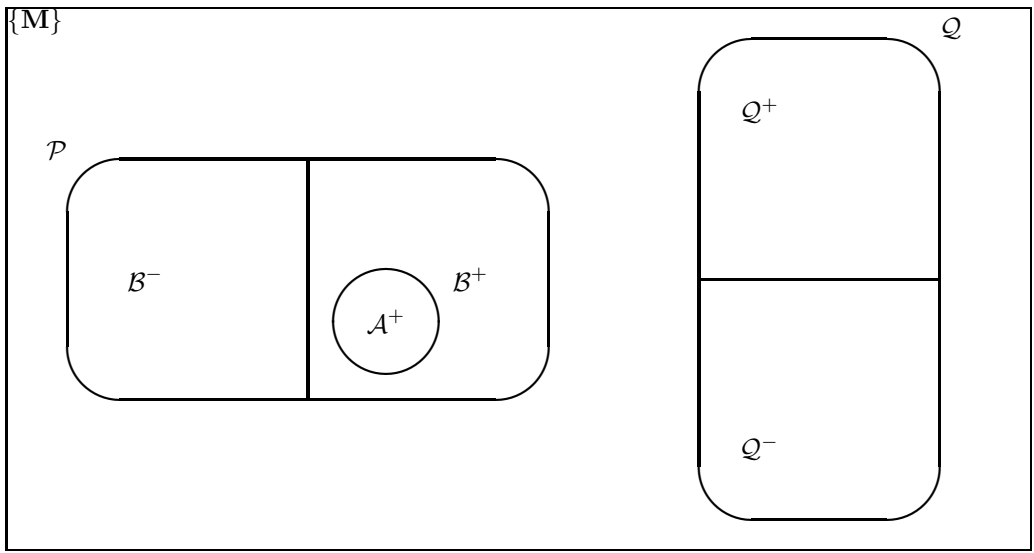

Hence the weaker orthomodularity condition is satisfied by
propositions about the 4-geon manifolds.  Quantum mechanics (as
represented on a complex Hilbert space) is a realization of
non-distributive proposition systems which satisfy
Equation~\ref{eq:quan}, and is believed to be unique as a
representation on a vector space. For a review and further references
on the relation between non-distributive proposition systems, quantum
mechanics and complex Hilbert spaces see
\cite[Chapters~21,22]{beltrametti_cassinelli}.

\section{CONSTRUCTION OF A MODULAR LATTICE}
By considering the measurements of the \x\ and \y\ components of spin
of a 4-geon with spin-half it is possible to construct a modular
lattice of propositions. It has been reported by Friedman and
Sorkin\cite{friedman_sorkin} that manifolds with the transformation
properties of a spinor can be constructed. For the construction which
follows, we require the 4-geon to have more than one possible outcome
from a Stern-Gerlach apparatus. We will consider two possible outcomes
($>0, \leq 0$); the exact spectrum, whether it is finite or infinite,
continuous or discrete is not important. The choice of \x\ and \y-spin
and the restriction to two outcomes is made to give a simple model of
the spin for a spin-half particle; momentum and position could equally
well have been used.

The relationship between orthomodular lattices and complex Hilbert
spaces described in References\cite{beltrametti_cassinelli,holland_s},
means that once we have constructed an orthomodular lattice of
propositions we can apply the internal symmetries and the symmetries
of space-time in the usual way\cite[Chapter~3]{ballentine} to
determine the form of the operators and the eigenvalues for spin,
momentum \etc\ The fact that a spin-half particle has two possible
values for the \x, \y\ or $z$ component of spin need not be assumed.

\begin{figure}[p]

\setlength{\unitlength}{.01mm}

\begin{picture}(12876,5452)(-599,-5191) 

\thicklines
\put(600,-2161){\circle*{336}}
\put(1800,-361){\line( 0,-1){1575}}
\put(3000,-361){\line( 0,-1){1575}}
\put(1800,-2380){\line( 0,-1){1575}}
\put(3000,-2380){\line( 0,-1){1650}}
\put(4200,-2761){\framebox(2100,1200){Filter}}

\thinlines
\put(5251,-2911){\vector( 1,-1){1050}}
\put(826,-1936){\vector( 3, 1){675}}
\put(826,-2311){\vector( 2,-1){660}}
\put(676,-2386){\vector( 1,-1){825}}
\put(676,-1936){\vector( 1, 1){825}}
\put(600,-1861){\vector( 0, 1){900}}
\put(600,-2461){\vector( 0,-1){900}}
\put(900,-2161){\vector( 1, 0){3000}}
\put(6376,-2161){\vector( 1, 0){3000}}
\put(300,-3736){\makebox(0,0)[lb]{Source}}
\put(2400,-4411){\makebox(0,0)[cb]{Collimator}} 
\end{picture}
\caption{The boundary conditions imposed by state-preparation}
\label{fig:init}
\end{figure}
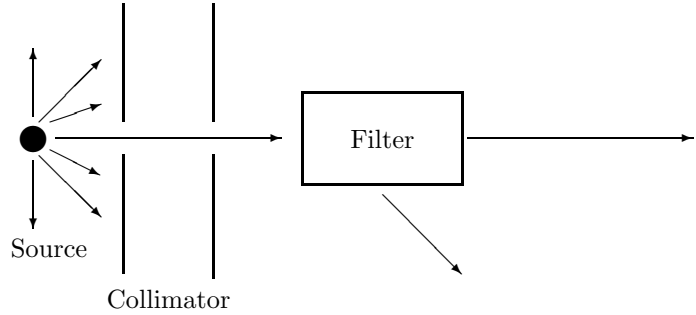

The set of all possible 4-geon manifolds, $\tilde{\MM}$, is not very
useful, since it includes manifolds compatible and incompatible with
every experimental arrangement. Let us constrain the possible
manifolds by setting up the state preparation apparatus as depicted in
Figure~\ref{fig:init}. By Axiom \ref{ax:state}, the apparatus imposes
boundary conditions which limit the set of relevant manifolds to $\MM
\subset \tilde{\MM}$, \ie\ to those 4-geons compatible with the
apparatus of Figure~\ref{fig:init}.
  
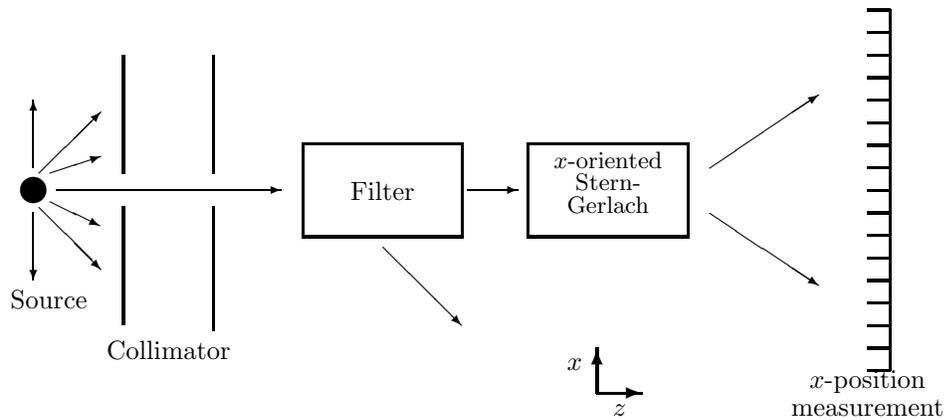
\begin{figure}[p]

\setlength{\unitlength}{.01mm}

\begin{picture}(12876,5452)(-599,-5191) 

\thicklines
\put(12000,239){\line( 0,-1){4800}}
\put(11700,239){\line( 1, 0){300}}
\put(11700,-61){\line( 1, 0){300}}
\put(11700,-361){\line( 1, 0){300}}
\put(11700,-661){\line( 1, 0){300}}
\put(11700,-961){\line( 1, 0){300}}
\put(11700,-1261){\line( 1, 0){300}}
\put(11700,-1561){\line( 1, 0){300}}
\put(11700,-1861){\line( 1, 0){300}}
\put(11700,-2161){\line( 1, 0){300}}
\put(11700,-2461){\line( 1, 0){300}}
\put(11700,-2761){\line( 1, 0){300}}
\put(11700,-3361){\line( 1, 0){300}}
\put(11700,-3661){\line( 1, 0){300}}
\put(11700,-3961){\line( 1, 0){300}}
\put(11700,-4261){\line( 1, 0){300}}
\put(11700,-4561){\line( 1, 0){300}}
\put(11700,-3061){\line( 1, 0){300}}
\put(11700,-4861){\makebox(0,0)[b]{$x$-position}}
\put(11700,-5161){\makebox(0,0)[b]{measurement}}
\put(7200,-2761){\framebox(2100,1200){}}
\put(8250,-1861){\makebox(0,0)[b]{\small $x$-oriented}}
\put(8250,-2161){\makebox(0,0)[b]{\small Stern-}}
\put(8250,-2461){\makebox(0,0)[b]{\small Gerlach}}
\put(600,-2161){\circle*{336}}
\put(1800,-361){\line( 0,-1){1575}}
\put(3000,-361){\line( 0,-1){1575}}
\put(1800,-2380){\line( 0,-1){1575}}
\put(3000,-2380){\line( 0,-1){1650}}
\put(4200,-2761){\framebox(2100,1200){Filter}}

\put(8100,-4861){\vector( 0, 1){600}}
\put(8100,-4861){\vector( 1, 0){600}}

\thinlines
\put(5251,-2911){\vector( 1,-1){1050}}
\put(826,-1936){\vector( 3, 1){675}}
\put(826,-2311){\vector( 2,-1){660}}
\put(676,-2386){\vector( 1,-1){825}}
\put(676,-1936){\vector( 1, 1){825}}
\put(600,-1861){\vector( 0, 1){900}}
\put(600,-2461){\vector( 0,-1){900}}
\put(900,-2161){\vector( 1, 0){3000}}
\put(9600,-2461){\vector( 3,-2){1453.846}}
\put(9600,-1861){\vector( 3, 2){1453.846}}
\put(6376,-2161){\vector( 1, 0){675}}
\put(300,-3736){\makebox(0,0)[lb]{Source}}
\put(2400,-4411){\makebox(0,0)[cb]{Collimator}}
\put(8400,-5161){\makebox(0,0)[b]{$z$}}
\put(7800,-4561){\makebox(0,0)[b]{$x$}}
\end{picture}
\caption{The boundary conditions imposed by state-preparation and an
$x$-spin measurement}
\label{fig:xmeas}
\end{figure}

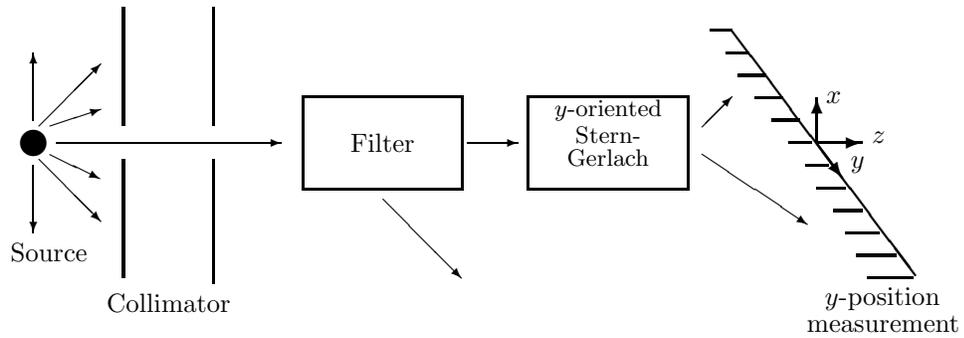
\begin{figure}[p]
\setlength{\unitlength}{.01mm}

\begin{picture}(12876,5452)(-599,-5191) 
\thicklines
\put(7200,-2761){\framebox(2100,1200){}}
\put(600,-2161){\circle*{336}}
\put(1800,-361){\line( 0,-1){1575}}
\put(3000,-361){\line( 0,-1){1575}}
\put(1800,-2380){\line( 0,-1){1575}}
\put(3000,-2380){\line( 0,-1){1650}}
\put(4200,-2761){\framebox(2100,1200){Filter}}

\put(11700,-3961){\line( 1, 0){600}}
\put(10200,-1561){\line( 1, 0){375}}
\put(10426,-1861){\line( 1, 0){375}}
\put(9889,-656){\line( 3,-4){2448}}
\put(11551,-3661){\line( 1, 0){525}}
\put(11400,-3361){\line( 1, 0){450}}
\put(11251,-3061){\line( 1, 0){375}}
\put(11026,-2761){\line( 1, 0){375}}
\put(10876,-2461){\line( 1, 0){300}}
\put(10651,-2161){\line( 1, 0){300}}
\put(9600,-661){\line( 1, 0){300}}
\put(9826,-961){\line( 1, 0){300}}
\put(9976,-1261){\line( 1, 0){375}}

\put(11026,-2161){\vector( 0, 1){600}}
\put(11026,-2161){\vector( 1, 0){600}}
\put(11026,-2176){\vector( 3,-4){340}}

\thinlines
\put(5251,-2911){\vector( 1,-1){1050}}
\put(826,-1936){\vector( 3, 1){675}}
\put(826,-2311){\vector( 2,-1){660}}
\put(676,-2386){\vector( 1,-1){825}}
\put(676,-1936){\vector( 1, 1){825}}
\put(600,-1861){\vector( 0, 1){900}}
\put(600,-2461){\vector( 0,-1){900}}
\put(900,-2161){\vector( 1, 0){3000}}
\put(6376,-2161){\vector( 1, 0){675}}

\put(9500,-1951){\vector( 1, 1){400}}
\put(9500,-2311){\vector( 3,-2){1400}}

\put(300,-3736){\makebox(0,0)[lb]{Source}}
\put(2400,-4411){\makebox(0,0)[cb]{Collimator}}

\put(7200,-2761){\framebox(2100,1200){}}
\put(8250,-1861){\makebox(0,0)[b]{\small $y$-oriented}}
\put(8250,-2161){\makebox(0,0)[b]{\small Stern-}}
\put(8250,-2461){\makebox(0,0)[b]{\small Gerlach}}

\put(11851,-2161){\makebox(0,0)[b]{$z$}}
\put(11251,-1636){\makebox(0,0)[b]{$x$}}
\put(11476,-2566){\makebox(0,0)[lb]{$y$}}
\put(11896,-4396){\makebox(0,0)[b]{$y$-position}}
\put(11911,-4681){\makebox(0,0)[b]{measurement}}

\end{picture}
\caption{The boundary conditions imposed by state-preparation and a
$y$-spin measurement}
\label{fig:ymeas}
\end{figure}

Next we can set up a Stern-Gerlach apparatus aligned with the \x-axis,
followed by an \x-position detector which here gives a value for the
spin (see Figure~\ref{fig:xmeas}). By Axiom \ref{ax:particle}, the
particle will certainly be detected at one position and only one
position. We denote by $\XX$ the 4-geon manifolds consistent with the
state-preparation and the \x-oriented Stern-Gerlach equipment. Clearly
$\XX \subseteq \MM$; because of the 4-geon postulate we can have $\XX
\neq \MM$. Of all the manifolds in $\XX$, some will correspond to $x >
0$, and the remainder to $x \leq 0$; these will be denoted $\XX^+$ and
$\XX^-$, respectively. Note that the same measurement apparatus
determines $x > 0$ and $x \leq 0$; therefore $\XX = \XX^+ \cup \XX^-$.

A \y-axis measurement may be made in a similar way (see
Figure~\ref{fig:ymeas}) which defines subsets $\YY$, $\YY^+$ and
$\YY^-$ of $\MM$. An \x\ and \y-oriented Stern-Gerlach apparatus
clearly cannot both be set in the {\em same} place at the {\em same}
time; they are incompatible, and by Axiom~\ref{ax:meas} the boundary
conditions which they set are incompatible. Hence $\YY$\ and $\XX$\
are disjoint subsets of $\MM$.

\subsection{The Propositions}
The propositions are the equivalence classes of outcomes of yes/no
experiments, two experiments being equivalent if there is no state
preparation that can distinguish them. Four non-trivial propositions,
$p$,$q$,$r$ and $s$, can be stated about the \x\ and \y-spin of 4-geon
manifolds, \M. They listed in Table~\ref{table:prop}, together with
the subsets of manifolds in the equivalence class and the experimental
results which they relate to.

In addition, there are the two trivial propositions $\To$ and
$\Ti$. Axiom~\ref{ax:particle} ensures that there exists at least one
4-geon manifold consistent with any measurement ($\exists \M \in
\XX$). Equivalently, given the state-preparation and measurement
boundary conditions then $\M \in \XX$. The trivial propositions, $\Ti$
(which is always true) and $\To$ (which is always false), correspond
to this Axiom and its converse:
\begin{equation}
\begin{array}{lcl}
\begin{array}[t]{rcl}
\To &\equiv&  \M \in \emptyset \nonumber \\
    &\equiv&  \XX =  \emptyset \nonumber \\
    &\equiv&  \YY =  \emptyset \nonumber 
\end{array} &
\hspace{10mm}&
\begin{array}[t]{rcl}
\Ti &\equiv&  \M \in \XX \mbox{\ for an \x-spin measurement}\nonumber \\
    &\equiv&  \M \in \YY \mbox{\ for a \y-spin measurement}\nonumber
\end{array} 
\end{array}
\end{equation}
The fact that the trivial propositions have more than one
interpretation is common to classical mechanics. For example, the
propositions that {\em the momentum is a real number} and that {\em
the position is a real number} are both always true for a classical
object. What is non-classical here is that these two physical
descriptions correspond to two different (and disjoint) sets of
possible results. Classically the measurements are different ways of
partitioning the common set defined by the initial conditions
alone. Here the measurements define two different sets, but the
propositions are identical because the sets give the same information.
\begin{table}[p]
\begin{tabular}{|lll|}\hline
Proposition & Manifolds & Measurement\\ \hline
$\To$& $\emptyset$& Always False \\
$p$    & $\M \in \XX^+$ & The \x-Spin is measured to be $ > 0$ \\
$q$    & $\M \in \XX^-$ & The \x-Spin is measured to be $\leq 0$ \\
$r$    & $\M \in \YY^+$ & The \y-Spin is measured to be $ > 0$ \\
$s$    & $\M \in \YY^-$ & The \y-Spin is measured to be $\leq 0$ \\
$\Ti$& $\M \in \XX$   & The \x-Spin is measurable\\
$\Ti$& $\M \in \YY$   & The \y-Spin is measurable\\ \hline
\end{tabular}
\caption{The propositions and sets of manifolds of the spin-half system}
\label{table:prop}
\end{table}

\subsection{Partial Ordering}
The ordering relation for two propositions, $a$ and $b$, is $a \leq b$
which means that $a$ true implies that $b$ is
true. For the spin-half system the partial ordering is almost trivial:
\begin{equation}
\To  \leq p  \leq \Ti, \hspace{10mm} 
\To  \leq q  \leq \Ti, \hspace{10mm}
\To  \leq r  \leq \Ti, \hspace{10mm}
\To  \leq s  \leq \Ti
\end{equation}
In this case there can be no ordering between $p$ and $r$ \etc\ when
they are in different directions, because $\XX$ and $\YY$ are disjoint
(and can clearly be distinguished by some state preparations) and so a
manifold cannot be in both. The propositions of the system therefore
form a poset (partially ordered set). Generally, the ordering relation
can only be applied to propositions if there exists at least one
experimental arrangement which evaluates both of them together.
 
\subsection{Meet and Join}
The meet of two propositions, $a\wedge b$, is the largest proposition,
the truth of which implies that both $a$ and $b$ are true. For any
poset it follows that: $a \wedge a = a$, $a \wedge \Ti = a$ and $a
\wedge \To = \To$. For this system we have in addition:
\begin{equation}
a\wedge b = \To ,\hspace{10mm}\hbox{$\forall a \neq b$} 
\end{equation}
For a 4-geon manifold, \M, to be in the meet of $p$ and $r$, it would
have to be in $\XX^+$ and $\YY^+$ which is not possible; the solution
set is therefore the empty set which corresponds to \To. Membership of
the subsets $\XX^+$ and $\YY^+$ corresponds to physically
distinguishable statements about the state preparation so the
equivalence relation does not affect this conclusion.

The join of two propositions, $a\vee b$, is the smallest proposition
which  is
true whenever either $a$ or $b$ is true. For any poset it
follows that: $a \vee a = a$, $a \vee
\To = a$ and $a \vee \Ti = \Ti$. For this system we have in addition:
\begin{equation}
a \vee b = \Ti \hspace{10mm}\hbox{$\forall a \neq b$} 
\end{equation}
In this small system there is no other acceptable choice for $p \vee
r$ \etc\ 

\subsection{Orthocomplementation}
As in classical mechanics we consider the orthocomplementation
$a^\perp$ of a proposition $a$ by taking the set-theoretic complement
with respect to all possible outcomes of the same experiment. We
define the complements of our system in Table~\ref{table:comp}

\begin{table}[p]
\begin{tabular}{|ll|}\hline
Complement & Manifolds \\ \hline
$\To^\perp = \Ti$&  Always True \\
$p^\perp = q$   & $\M \in (\XX\setminus\XX^+ \equiv \XX^-)$ \\
$q^\perp  =p$  & $\M \in (\XX\setminus\XX^- \equiv \XX^+)$   \\
$r^\perp  =s$  & $\M \in (\YY\setminus\YY^+\equiv \YY^-)$  \\
$s^\perp  =r$  & $\M \in (\YY\setminus\YY^-\equiv \YY^+)$ \\
$\Ti^\perp = \To$  & Always False \\ \hline
\end{tabular}
\caption{The complements of the propositions of the spin-half system}
\label{table:comp}
\end{table}

From Table~\ref{table:comp} and Table~\ref{table:prop}, it is clear
that the required properties of orthocomplementation are satisfied:
\begin{enumerate}
\item $(a^\perp)^\perp = a$ 
\item $a \vee a^\perp = \Ti$ and $a \wedge a^\perp = \To$
\item $a \leq b \Rightarrow b^\perp \leq a^\perp$
\end{enumerate} 
The first two follow directly from set theory, while the third
only applies in the cases: $a<\Ti$ or $\To <a$, because of the
simple structure of this poset.  

The definition given satisfies DeMorgan's laws:
\begin{eqnarray}
(a_1 \wedge a_2 )^\perp &=& a_1^\perp \vee a_2^\perp \\
(a_1 \vee a_2 )^\perp &=& a_1^\perp \wedge a_2^\perp \\
\end{eqnarray}
Thus we have an orthocomplemented poset. DeMorgan's Laws can be used
to define the join of two incompatible propositions in terms of the
meet and orthocomplementation \eg:
\begin{equation}
 p \vee r = (q\wedge s)^\perp = \To^\perp = \Ti
\end{equation}

\subsection{Lattice}
A lattice is a poset where the meet and join always exist. The meet
and join of any two elements of this system always exist, these being
\To\ and \Ti\, respectively, for any two different
propositions. Table~\ref{table:meet} shows the meet and join for all
the propositions.
\begin{table}[p]
\begin{tabular}{ccc}
\begin{tabular}{|l|cccccc|}\hline
$\wedge$&$\To$&  $p$&  $q$&  $r$&  $s$&$\Ti$ \\ \hline
$\To$   &$\To$&$\To$&$\To$&$\To$&$\To$&$\To$ \\
$p$     &$\To$&$p$  &$\To$&$\To$&$\To$&$\To$ \\
$q$     &$\To$&$\To$&$q$  &$\To$&$\To$&$\To$ \\
$r$     &$\To$&$\To$&$\To$&$r$  &$\To$&$\To$ \\
$s $    &$\To$&$\To$&$\To$&$\To$&$s$  &$\To$ \\
$\Ti$   &$\To$&$\To$&$\To$&$\To$&$\To$&$\Ti$ \\ \hline
\end{tabular} &
\hspace{10mm} &
\begin{tabular}{|l|cccccc|}\hline
$\vee$ &$\To$& $p$ & $q$ & $r$ & $s$ &$\Ti$ \\ \hline
$\To$  &$\To$&$\Ti$&$\Ti$&$\Ti$&$\Ti$&$\Ti$ \\
$p$    &$\Ti$&$p $ &$\Ti$&$\Ti$&$\Ti$&$\Ti$   \\
$q$    &$\Ti$&$\Ti$&$q$  &$\Ti$&$\Ti$&$\Ti$   \\
$r$    &$\Ti$&$\Ti$&$\Ti$&$r$  &$\Ti$&$\Ti$   \\
$s$    &$\Ti$&$\Ti$&$\Ti$&$\Ti$&$s$  &$\Ti$  \\
$\To$  &$\Ti$&$\Ti$&$\Ti$&$\Ti$&$\Ti$&$\Ti$ \\ \hline
\end{tabular}\\
\end{tabular}
\caption{The meets  and joins of the propositions of the spin-half system}
\label{table:meet}
\end{table}

The poset is thus seen to be an orthocomplemented {\em Lattice}.

\subsection{Orthomodularity} 
The orthomodularity condition:
\begin{equation}
a \leq b \Rightarrow b = a \vee (b \wedge a^\perp)
\label{eq:orthomodular}
\end{equation}
is satisfied by the simple spin-half poset, as can be seen by
considering each case, $\forall a \in\{p,q,r,s\}$:
\begin{eqnarray} 
\To \leq a   &\Rightarrow& a = \To \vee (a \wedge \Ti) \\
a \leq \Ti   &\Rightarrow& \Ti = a \vee (\Ti \wedge a^\perp) \\
a \leq a     &\Rightarrow& a = a \vee (a \wedge a^\perp) \\
\To \leq \Ti &\Rightarrow& \Ti = \To \vee (\Ti \wedge \Ti) 
\end{eqnarray}

\subsection{Modularity}
That this lattice is modular can be seen by examining it case by
case. The failure of the modularity law, as required for a strictly
orthomodular lattice, will only occur for systems with an infinite
spectra \cite[Page~220]{jauch}.

\subsection{Distributivity}
A simple counterexample suffices to show that the distributive rule
fails for propositions about different directions:
\begin{equation}
p \wedge (r \vee r^\perp) \neq (p \wedge r) \vee (p \wedge r^\perp)
\end{equation}
the LHS is $p \wedge \Ti = p$, while the RHS is $\To \vee \To = \To$;
thus $p$ and $r$ are not compatible. The result can be checked from
Table~\ref{table:meet} of meets and joins or by noting that the
subsets $\XX^+,\YY^+,\YY^-$ corresponding to the propositions $p$,$r$
and $s$, respectively, are all disjoint and not related by the
equivalence relation.

\subsection{Atomicity}
An atom is a proposition, different from $\To$, which does not have
any smaller proposition. The propositions $p,q,r,s$ are clearly the
atoms.

\subsection{Covering Property}
We say that $a$ covers $b$ if $a > b$, and $a \ge c \ge b$ implies
either $c =a$ or $c = b$. A lattice has the covering property if the
join of any element, $a$, with an atom not contained in $a$ covers
$a$. Clearly $\forall a,b \in \{p,q,r,s\}$:
\begin{eqnarray}
\To \vee a &=& a \mbox{\hspace{5mm} which covers $\To$} \\
a \vee b &=& \Ti\mbox{\hspace{5mm} which covers $a$}
\end{eqnarray}
This establishes that the system has the covering property.

Starting with propositions about sets of manifolds in classical
general relativity, we have constructed a non-distributive,
orthomodular lattice, which is atomic and has the covering
property. The significance is not just that such a lattice is a
feature of quantum mechanics, but that it is {\em the} distinguishing
feature of quantum mechanics. It has previously been thought that a
non-distributive lattice of propositions could never be constructed
from a classical theory and hence that no classical explanation of
quantum mechanics was possible; this is shown to be false. The present
work gives a classical explanation for the origin of quantum mechanics
and because it is based on the accepted theory of general relativity,
it offers the most economical interpretation.

\section{CONCLUSIONS}
By modelling particles as 4-geons in general relativity (rather than
evolving 3-manifolds), features characteristic of quantum mechanics
can be derived. This work therefore offers a novel possibility for a
classical basis for quantum mechanics, and in doing so offers a way to
reconcile general relativity and quantum mechanics.  Some of the
implications and unresolved issues are:
\begin{enumerate}
\item Time is an asymptotic approximation as expected by
workers in quantum gravity\cite{isham}.

\item The theory does not exclude classical objects. Measurements of a
3-geon, if one existed, could not satisfy the logic of quantum
mechanics. Gravitational waves are also described as evolving
3-manifolds and, although there may be some problems defining a global
hypersurface \cite{penrose}, they do not have the topological structure to
exhibit the measurement-dependent effects characteristic of quantum
mechanics.

\item It follows from the previous comment that there is no
graviton. Potentially, this is a testable prediction of the theory.

\item Even in the asymptotic region, the metric associated with our
4-geon model of a single particle is not well-defined by state
preparation alone, since each manifold consistent with the state
preparation ($\M \in \MM$) can have different asymptotic
properties. This is an almost inevitable consequence of reconciling
quantum mechanics and general relativity\cite{page_geilker}. However, the
present perspective on the origin of quantum mechanics accounts for
the lack of a well-defined metric as being due to incompleteness of
boundary conditions imposed by state preparation, rather than as an
inherent feature of the gravitational field.

\item Non-local effects - as exemplified by the EPR experiments - can
be explained by theories with non-trivial
topologies\cite[Page~481]{holland}, since with CTCs there can exist
causal routes from one arm of the experiment to the other. Quantum
mechanics itself requires only a failure of weak-causality
(statistical correlations of a non-local character between spacelike
separated events\cite{ballentine_jarrett}); with CTCs as an essential
part of the structure of an elementary particle however, it is not
clear why a failure of strong-causality (communication between
spacelike separated events) is not apparent.
 
\item Like all theories of quantum gravity and interpretations of
quantum mechanics, this work is speculative. The theory can only be
proven if exact
solutions to Einstein's equations with the required properties
are found. Considering the difficulty of finding exact solutions with
non-trivial topology, a more practical way of confirming these ideas is
to examine the predictions, the first of which is the absence of a
graviton.

\end{enumerate}

It would indeed be ironic if the interpretation quantum theory with
which Einstein was so dissatisfied could be seen to be a consequence
of his general theory of relativity.

\section{Acknowledgements}
I would like to thank Dr Hyland and Professor Isham for helpful
discussions. This work was supported by the University of Warwick


\end{document}